\documentclass[aps,prb,reprint,superscriptaddress]{revtex4-2}
\usepackage[utf8]{inputenc}
\usepackage{graphicx}
\usepackage[table,xcdraw]{xcolor}
\usepackage[american]{babel}
\usepackage{hyperref}
\usepackage[normalem]{ulem}

\usepackage{amsmath}
\usepackage{microtype}
\makeatletter
\newcommand{\equalcontribmark}{%
  \gdef\comma@space{\textsuperscript{,\,*,\,}}%
}
\makeatother

\begin{document}
\preprint{APS/123-QED}
\title{Variable Charge State, Magnetic Excitations, and Kondo Effect of Sm/g/Ir(111)}

\author{Shixuan~Shan}
\thanks{These authors contributed equally to this work.}
\affiliation{Institute of Physics, École Polytechnique Fédérale de Lausanne (EPFL), CH-1015 Lausanne, Switzerland}

\author{Tamara~de~Ara}
\thanks{These authors contributed equally to this work.}
\affiliation{Institute of Physics, École Polytechnique Fédérale de Lausanne (EPFL), CH-1015 Lausanne, Switzerland}

\author{Lina~Liu\equalcontribmark}
\altaffiliation[Present address: ]{%
Center for Spintronics and Quantum Systems,
State Key Laboratory for Mechanical Behavior of Materials,
School of Materials Science and Engineering,
Xi'an Jiaotong University,
Xi'an 710049, China}
\affiliation{Institute of Physics, École Polytechnique Fédérale de Lausanne (EPFL), CH-1015 Lausanne, Switzerland}

\author{Zhipeng~Wang}
\affiliation{Department of Chemistry, Graduate School of Science, Tohoku University, Aramaki-Aza-Aoba, Aoba-Ku, Sendai 980-8578, Japan}

\author{Marina~Pivetta}
\author{François~Patthey}
\affiliation{Institute of Physics, École Polytechnique Fédérale de Lausanne (EPFL), CH-1015 Lausanne, Switzerland}

\author{Tadahiro Komeda}
\affiliation{Department of Chemistry, Graduate School of Science, Tohoku University, Aramaki-Aza-Aoba, Aoba-Ku, Sendai 980-8578, Japan}
\affiliation{Institute of Multidisciplinary Research for Advanced Materials (IMRAM), Tohoku University, 2-1-1 Katahira, Aoba-ku, Sendai 980-8577, Japan}
\affiliation{Center for Spintronics Research Network, Tohoku University, 2-1-1 Katahira, Aoba-ku, Sendai 980-8577, Japan}

\author{Daria~Kývala}
\author{Jindřich~Kolorenč}
\affiliation{Institute of Physics (FZU), Czech Academy of Sciences, Na Slovance 2, 182 00 Praha, Czech Republic}

\author{Harald~Brune}
\email{harald.brune@epfl.ch}
\affiliation{Institute of Physics, École Polytechnique Fédérale de Lausanne (EPFL), CH-1015 Lausanne, Switzerland}

\date{\today}

\begin{abstract}
Using low-temperature scanning tunneling microscopy we investigate the charge state, magnetic excitations, and Kondo features of individual Sm adatoms on graphene/Ir(111). Depending on the number and distance of their neighbors, Sm atoms can be in two discrete charge states. At certain distances, a reversible transition between these two states is induced by the electric field of the STM tip leading to concentric charge rings in the images. Only atoms in one of the two charge states exhibit magnetic excitations in d$I$/d$V$ spectra. Two such excitations are located at 35~meV and 54~meV and related to transitions from the $J = 1/2$ ground state doublet to the first crystal field split $J = 3/2$ multiplet. Together with the intra-atomic exchange excitations at higher energy, these observations indicate that Sm transfers one $6s$ electron to the substrate while it retains its gas-phase $4f$ filling. New for lanthanide adatoms, we observe a Kondo resonance. The Zeeman splitting of the Kondo peak reveals that Sm retains its large gas-phase $g$-factor. Comparison of d$I$/d$V$ spectra to cotunneling theory yields the crystal field acting on the $4f$ shell and, consequently, on the $J = 3/2$ quadruplet, confirms Sm$^+$ as the ground state, and identifies Sm$^{2+}$ as being energetically close, thereby rationalizing our observation of variable charge states.

\end{abstract}
\maketitle

\section{Introduction}
A single atom magnet (SAM) is an individual surface adsorbed atom that exhibits magnetic hysteresis, thereby enabling magnetic information storage in the smallest unit of matter~\cite{don16, bal16}. SAMs can be placed as close as 1.2~nm and individually magnetized in any one of the two stable states~\cite{nat17, sin21}, translating to potential magnetic storage densities of 450 Tera bits/in$^2$. All SAMs known today are rare-earth atoms adsorbed either on ultra-thin decoupling layers grown on single-crystal metal surfaces~\cite{don16, bal16, sin21, sor23, piv25, per26} or on single crystal surfaces of wide band gap semiconductors~\cite{bel22}. 

While the 4$f$ filling of the respective rare-earth atom can be inferred from X-ray absorption spectroscopy (XAS), the occupation of the valence shells and the resulting charge state are more difficult, if not impossible, to measure with this technique~\cite{sin21orb}. However, it determines to a large extent the SAMs' magnetic properties, very similar to molecular magnets~\cite{gou19, dub19}. The valence spin, resulting from partially filled valence shells, is exchange coupled to the spin of the strongly localized $4f$ electrons. Together they form the total angular moment, its projections onto a symmetry direction determine the magnetic level scheme and the mechanisms of magnetization reversal~\cite{dub21, cur23}. Moreover, the spin-polarization of the valence shells enables access to the atoms' properties by means of scanning tunneling microscopy (STM), either by magnetic excitation spectroscopy~\cite{hei04}, magnetic d$I$/d$V$~\cite{mei08} and apparent height~\cite{nat17} contrast, or by electron spin resonance (ESR)~\cite{bau15, cza25}.

For systems with magnetic bi-stability on a time scale accessible to STM, the magnetic contrast can be taken as a degree of spin-polarization thereby providing an estimate of the filling of the valence orbitals~\cite{nat18, piv20, sin21}. For systems that don't exhibit this contrast, estimates can be derived from electrostatic interactions between the individual adatoms. For Sm atoms on a graphene monolayer on Ir(111), nucleation experiments have revealed a mutual repulsion that originates from electric dipole-dipole interactions~\cite{piv18}. Their strength translates to a charge transfer of $0.83 \pm 0.03$ electrons from Sm to the substrate. Hence to a good approximation the atoms are Sm$^+$ with an almost fully spin-polarized valence shell. In line with this, intra-atomic spin excitations between 4$f$ and valence spin can be observed as d$I$/d$V$-steps in STS~\cite{piv20}. This singly charged cation Sm$^+$ is of particular interest, for in its 4$f^6$ state it combines a large orbital moment $L = 3$ and a large Landé $g$-factor with a total angular moment of only $J = 1/2$~\cite{nist, mar78, cza25}, making it a potential qubit candidate, accessible to ESR-STM~\cite{rea24, cza25}.

Here we report that the charge state of individual Sm atoms on g/Ir(111) depends on their mutual distance and on the number of Sm neighbors. While isolated adatoms are in a charge state where they show no magnetic excitations, atoms closer to each other than 1.5~nm are in a bistable charge state giving rise to a concentric charge ring in STM images. These atoms always exhibit magnetic excitations. In addition to the already reported intra-atomic spin excitations at 160--170~meV~\cite{piv20}, we observe two prominent d$I$/d$V$ steps at lower energies. Both are associated with ${\bf S}_{4f}$ and {\bf L} going from the fully antiparallel ground state to partial canting, resulting in $J_{4f} = 1$ and $J = 3/2$~\cite{nist, cza25}. In a finite crystal field, this state has two non-degenerate $z$-projections, $J^{\rm tot}_{\rm z} = 3/2$ and $J^{\rm tot}_{\rm z} = 1/2$. The d$I$/d$V$-step at smallest $E$ corresponds to the first, and has been observed for the same atom on MgO~\cite{cza25}, while the hitherto not reported excitation at slightly higher $E$ promotes Sm$^+$ to the second $J$ projection. The atoms exhibit a Kondo peak with a Zeeman splitting that confirms their very large $g$-factor. 

\section{Experimental details}
The experiments were done with a homebuilt ultra-high vacuum (UHV) 0.4~Kelvin 8.0/0.8~Tesla STM~\cite{nat19, swe25}. The Ir(111) surface was cleaned by repeated sputter ($p_{\rm Ar^+} = 1.3 \times 10^{-7}$mbar, 300~K, 30~min) and annealing (1450~K, 5~min) cycles. Graphene was grown by temperature-programmed growth (TPG)~\cite{cor09}, since this method yields graphene islands, leaving bare metal areas for tip preparation. The sample was exposed to 90 Langmuir ethane at room temperature ($4 \times 10^{-7}$~mbar for 5~min). The chamber was then pumped down to UHV ($1 \times 10^{-10}$~mbar), followed by flash annealing to 1200~K for 60~s. Subsequently, the sample was cooled to room temperature and transferred to the STM being at 4~K. Sm atoms were deposited onto the sample in the STM from a thoroughly degassed high-purity rod (99.9~\%) using an $e$-beam evaporator. A current of $80-100$~nA on the flux monitor for 40~s yielded a Sm coverage of $\Theta_{\rm Sm} = 8 \times 10^{-3}$~monolayers (ML), where one ML is defined as one Sm atom per graphene unit cell. Opening the STM shutter during deposition led to a small temperature increase on the sample. However, its temperature was kept below 10~K in order to suppress diffusion of the Sm adatoms. After deposition, the sample was cooled to the base temperature of 0.4~K, where the majority of STM measurements have been performed. We used Pt–Ir tips. The tunnel voltage $V_{\rm t}$ was applied to the tip, with the sample on virtual ground via the $IV$-converter. d$I$/d$V$-spectra were acquired using Nanonis' built-in lock-in module, with a modulation frequency of 323~Hz and a modulation amplitude of 3~mV for spectra covering a wide voltage range and 1~mV for narrow voltage range spectra. Atom manipulation with the STM tip was achieved with $V_{\rm t} = 2$~mV and a tunnel current of $I_{\rm t} = 700$~pA with the feedback loop disabled. Unless explicitly mentioned, all experiments were performed under an out-of-plane magnetic field of 0.5~T, since this damped the occasional pinging noise of our cryostat.

\section{Results and discussion}
\subsection{Distance-dependent charge state}
\label{sec:charge_and_dist}
\begin{figure}[h!]
	\centering
	\includegraphics[width = \linewidth]{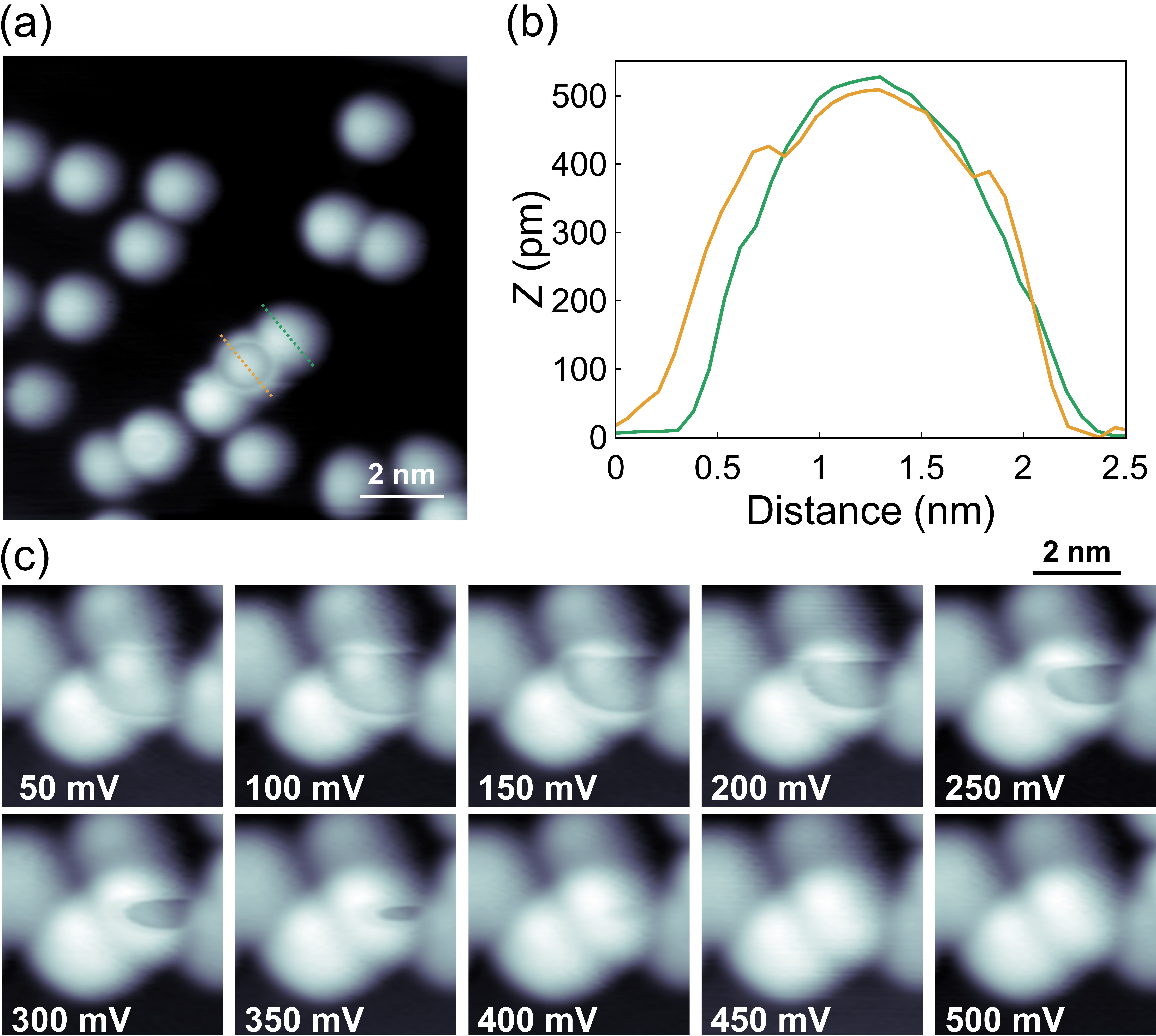}
	\caption{(a) STM image of individual Sm adatoms on a graphene island on Ir(111) ($\Theta_{\rm Sm} = 8 \times 10^{-3}$~ML, $V_{\rm t} =~100$~mV, $I_{\rm t} = 10$~pA). One atom shows a particular charge state manifested by a concentric ring, while all other atoms appear as regular smooth protrusions. (b) Profiles taken through a regular adatom (green) and through the one with a charge ring (yellow). (c) The charge ring changes radius as the tunnel voltage is varied, pointing to an electric field effect ($I_{\rm t} = 10$~pA).}
	\label{cha}
\end{figure}

For samples with very small Sm coverages, the majority of the Sm adatoms are imaged with STM as regular protrusions with the characteristic smooth Gaussian profile resulting from tip-atom convolution, see the green apparent height profile in Fig.~\ref{cha}(b) and the dashed line in the STM image Fig.~\ref{cha}(a) showing where this profile was taken. In contrast, atoms that have several close nearest neighbors may exhibit a concentric ring that is visible as kink in their profile shown in yellow. We attribute this ring to a reversible change of charge state of the Sm adatom occurring when the tip is at a certain distance from the center of the adatom. The radius of the charge ring changes with tunnel voltage, as illustrated for the central atom in Fig.~\ref{cha}(c), pointing to an effect induced by the electric field between the tip and sample. A reversible change in adsorption site can be excluded since this would be a lateral displacement by a discrete and constant amount along any of the 6 high-symmetry directions of the substrate, giving rise to discrete features, while we observe a perfect concentric ring. 

Atoms with different charge states adsorbed on thin insulating layers on metal surfaces are known to exhibit different STM appearance. Au$^-$/NaCl(100)/Cu(111) has a depletion ring around it, while Au$^0$ on the same surface doesn't, but appears with significantly larger height~\cite{gro09}. While these features are not altered by the measurement and are permanently there for the respective charge states, here we observe a dynamic and discrete change of the charge state, caused by the varying electric field in the tunnel junction as the tip scans over the adatom. The Sm atoms exhibiting this ring are very close to a transition between two discrete charge states, and this state reversibly switches when the tip apex is at a certain distance from the center of the adatom, giving rise to the ring. 

For the present system, it was inferred from self-assembly studies that electron transfer from Sm to the substrate takes place and manifests itself by a screened repulsive Coulomb potential (screening length $r_0 = 0.75$~nm and charge $q =  0.83 \pm 0.03\,e$) as well as from a Sm coverage-dependent shift of the Ir(111) surface state onset to higher binding energies~\cite{piv18}. This shift is non-linear with a much reduced slope for $\Theta_{\rm Sm} > 1 \times 10^{-2}$~ML. Hence once the atoms come closer than the mean distance $d = 2.5$~nm corresponding to that coverage, the dipole-dipole repulsion strongly reduces the charge transfer per atom. Thus the charge transfer per atom varies with interatomic distance, and the atoms with close neighbors must be between two discrete charge states, possibly below and above one electron transferred per atom, and the presence of the electric field of the tip toggles the atoms between the two states. 

\begin{figure}[h!]
	\centering
	\includegraphics[width = \linewidth]{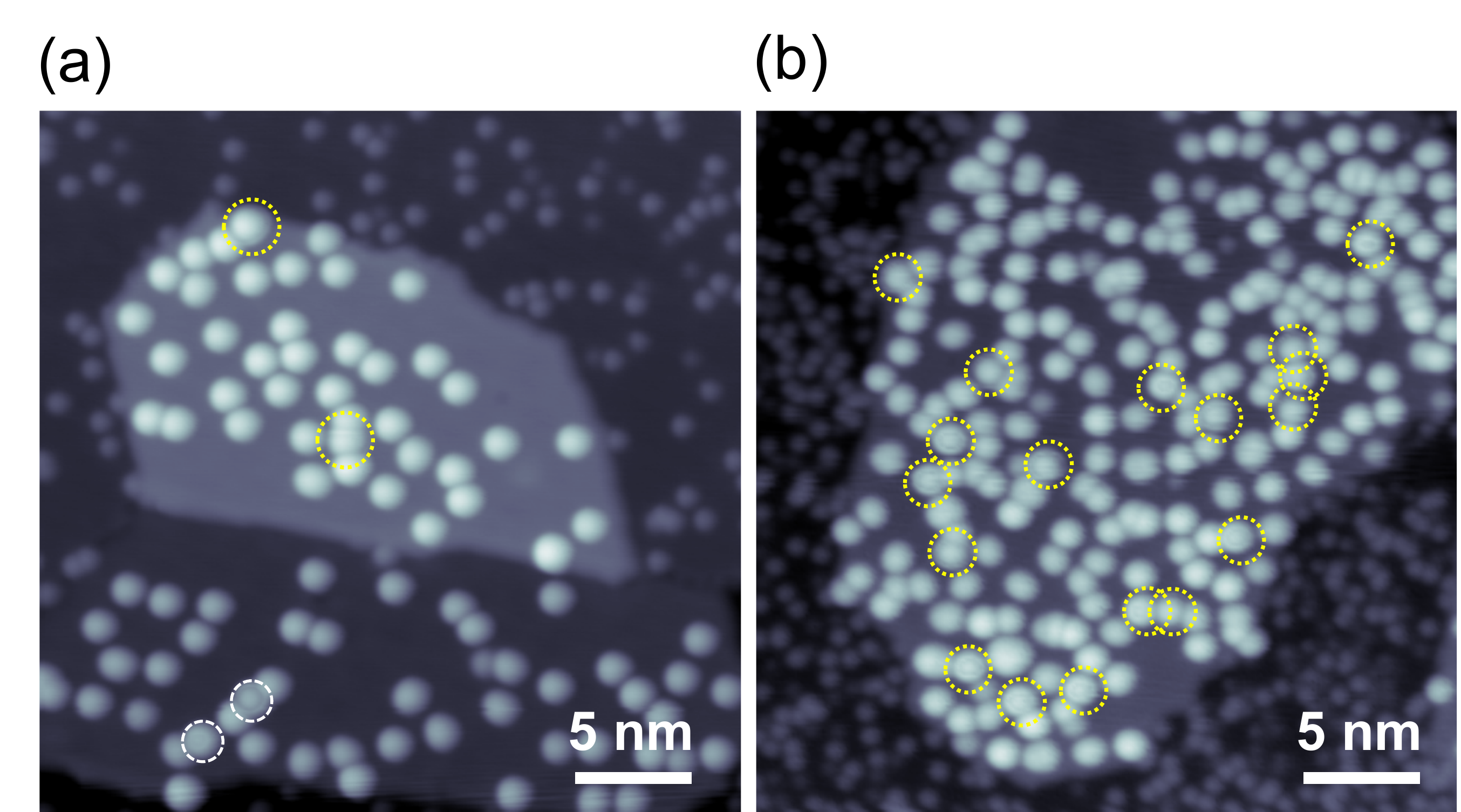}
	\caption{Coverage dependent abundance of adatoms with charge ring. (a) The upper half shows a graphene island surrounded by Ir(111) substrate, the lower part is a graphene island on the next lower atomic terrace. Atoms with charge ring are marked with dashed yellow circles on the upper island and with dashed white ones on the lower island ($\Theta_{\rm Sm} =~8~\times~10^{-3}$~ML, $V_{\rm t} = 100$~mV, $I_{\rm t} = 10$~pA). (b) Sample with 2.5 times higher Sm coverage. There is one graphene island surrounded by Ir(111). Atoms with charge ring are marked with dashed yellow circles ($\Theta_{\rm Sm} = 2 \times 10^{-2}$~ML, $V_{\rm t} = 100$~mV, $I_{\rm t} = 10$~pA).}
	\label{cov}
\end{figure}

Increasing the coverage of Sm atoms increases the fraction of atoms that have sufficiently close neighbors to exhibit the charge ring. While Fig.~\ref{cov}(a) displays only four atoms ($\sim$ 4.5 \% of the atoms on the graphene) with charge ring, Fig.~\ref{cov}(b) has 18 ($\sim$ 8.1 \%) on a comparable surface area at 2.5 times higher coverage.

We note that similar charge rings have been observed for individual Dy atoms on g/Ir(111) (see Fig. \ref{Dy}).

\subsection{Charge state and magnetic excitations}
\label{sec:charge_and_magexc}
A necessary condition for the appearance of magnetic excitations related to intra-atomic Heisenberg exchange interactions, is spin-polarization in the valence shells~\cite{piv20}. Hence, if an adatom shows the corresponding steps in d$I$/d$V$-curves, the valence shells must be partly filled, giving indirect information about the charge state. Therefore we investigated the correlation of charge rings and the appearance of d$I$/d$V$-steps. Atoms with charge rings always exhibit inelastic d$I$/d$V$-steps, the most prominent ones being at 160--170~mV and related to a transition from parallel to anti-parallel 4$f$ and 6$s$ spins~\cite{piv20}. However, on samples with high enough coverage, also some of the atoms without charge ring present these steps. Therefore the charge ring is a sufficient but not a necessary condition to exhibit magnetic d$I$/d$V$-steps.

\begin{figure}[b!]
	\centering
	\includegraphics[width = \linewidth]{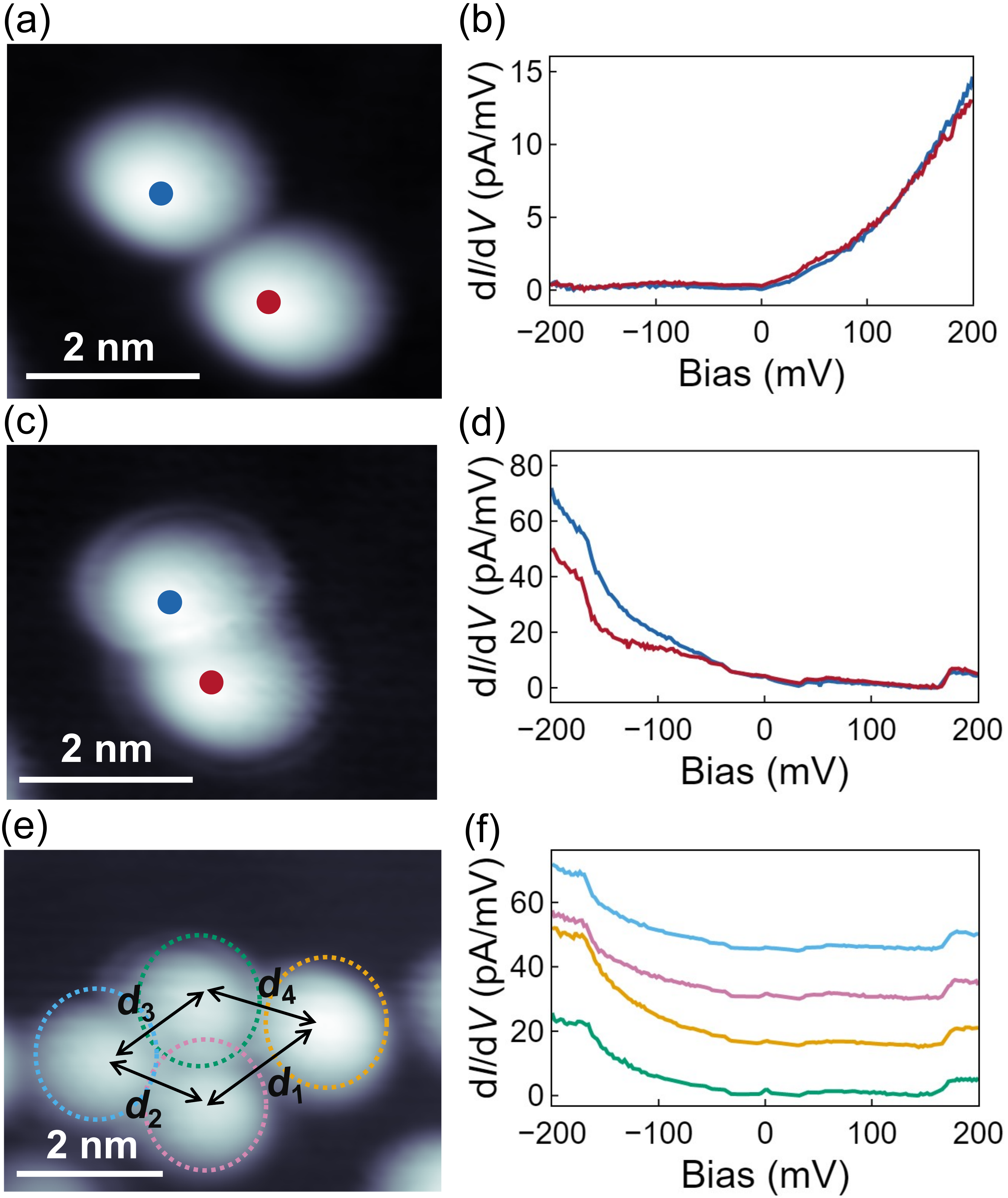}
	\caption{(a) and (c) STM images of two isolated Sm adatoms that are approached by atomic manipulation; their summits are marked in red and blue. (b) and (d) d$I$/d$V$-spectra of the two adatoms ($V_{\rm t} = 200$~mV, $I_{\rm t} = 400$~pA). Inelastic conductance steps appear at 160--170~mV once the atoms are approached. (e) An ensemble of four Sm adatoms with varying interatomic distances $d_1$--$d_4$ ($V_{\rm t} = 100$~mV, $I_{\rm t} = 10$~pA). (f) The corresponding d$I$/d$V$-curves display the inelastic conductance steps for each of the four Sm atoms. The spectra are vertically offset for clarity.}
	\label{man}
\end{figure}

We investigated the appearance of d$I$/d$V$-steps and of the charge rings as function of the interatomic distance by approaching two isolated Sm adatoms with atomic manipulation. At $d = 1.87 \pm 0.04$~nm neither of the Sm adatoms exhibits d$I$/d$V$ steps nor charge rings, see Figs.~\ref{man}(a) and (b).
Approaching them to $d = 0.98 \pm 0.07$~nm, d$I$/d$V$-steps and charge rings appear on both atoms, as can be seen from Figs.~\ref{man}(c) and (d). The critical distance below which d$I$/d$V$-steps and charge rings appear varies from 0.9 to 1.5~nm, depending on the total Sm coverage but also on the location of second and third neighbor Sm atoms. For example, this critical distance gets larger as more atoms are approached toward each other. Figure~\ref{man}(e) shows four atoms that were approached until they all showed magnetic d$I$/d$V$-steps, see Fig.~\ref{man}(f). The largest interatomic distance in this artificial Sm island is $1.68 \pm 0.04$~nm and thereby larger than the critical distance in the case of two isolated adatoms. Note that the energy of the magnetic excitations is independent of the interatomic distance within our experimental accuracy. 

\subsection{Adsorption site in the graphene moiré}
\begin{figure}[b!]
	\centering
	\includegraphics[width = \linewidth]{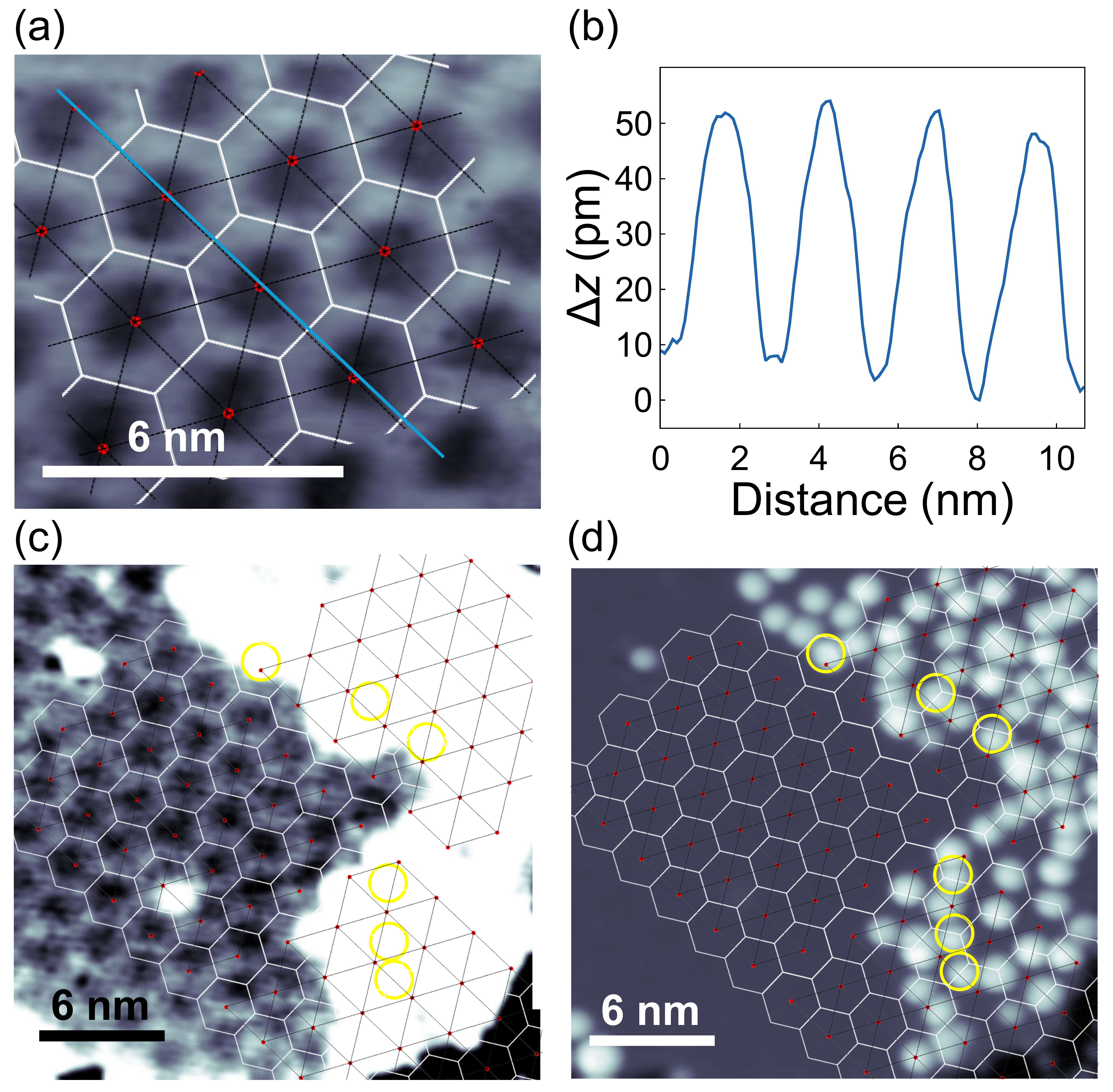}
	\caption{(a) Moiré pattern of g/Ir(111) marked as honeycomb (white) and hexagonal (black) lattice ($V_{\rm t} =~100$~mV, $I_{\rm t} =~10$~pA). (b) Profile along the blue line in (a). (c) and (d) Sample prepared to localize the Sm adatoms on the moiré unit cell. (c) Contrast shows the moiré on the left side, where Sm atoms have been removed. (d) Contrast shows the adatoms on the extrapolated moiré pattern. Yellow circles mark Sm adatoms with d$I$/d$V$ steps ($V_{\rm t} = 100$~mV, $I_{\rm t} = 10$~pA).}
	\label{moire}
\end{figure}

To accommodate the lattice mismatch with the Ir(111) substrate, graphene forms a moiré pattern with a $(9.32~\pm~0.15~\times~9.32~\pm 0.15)$ unit cell~\cite{ndi08}. The location of the Sm atoms within this unit cell could have an influence on the charge state and on the occurrence of magnetic d$I$/d$V$-steps. Figure~\ref{moire}(a) shows an STM image of the moiré pattern. 
The areas with the lowest apparent height are the stacking regions where graphene rings are centered atop Ir atoms~\cite{ndi08}. They are marked with red dots, surrounded by white hexagons and connected by dashed black lines, representing the moiré unit cells once as honeycombs and once as diamonds. The corresponding height profile in Fig.~\ref{moire}(b) shows a cut along the blue line revealing the moir\'{e} corrugation of $(40 \pm 3)$~pm, in good agreement with 45~pm reported from non-contact AFM experiments with CO terminated tips~\cite{bon14}, and with 35~pm found in DFT calculations~\cite{bus11}.

In order to localize the adatoms on the moiré cells we prepared the sample shown in Figs.~\ref{moire}(c) and (d). On the left-hand side of the atomic terrace we removed by atomic manipulation all the Sm adatoms, making the moiré pattern clearly apparent and enabling its extrapolation onto the right-hand side, where the Sm adatoms are still located on their initial adsorption sites. The same image is shown once with a gray scale where the moiré pattern can be seen (c) and once with a gray scale where the location of the Sm atoms on the extrapolated moiré pattern can be identified (d). The Sm atoms are randomly distributed over the moiré unit cells, as expected from statistical growth reported before for $T_{\rm dep} < 10$~K~\cite{piv18}. Yellow circles mark those adatoms for which d$I$/d$V$-steps such as in Fig.~\ref{man}(d) and (f) were observed. They are adsorbed on different stacking areas of the respective moiré unit cell, hence we exclude a strong influence of the location in the moiré on the adatom charge state.

\subsection{Magnetic excitations at low energy}
\label{sec:magexc}
\begin{figure}[t!]
\centering
\includegraphics[width = \linewidth]{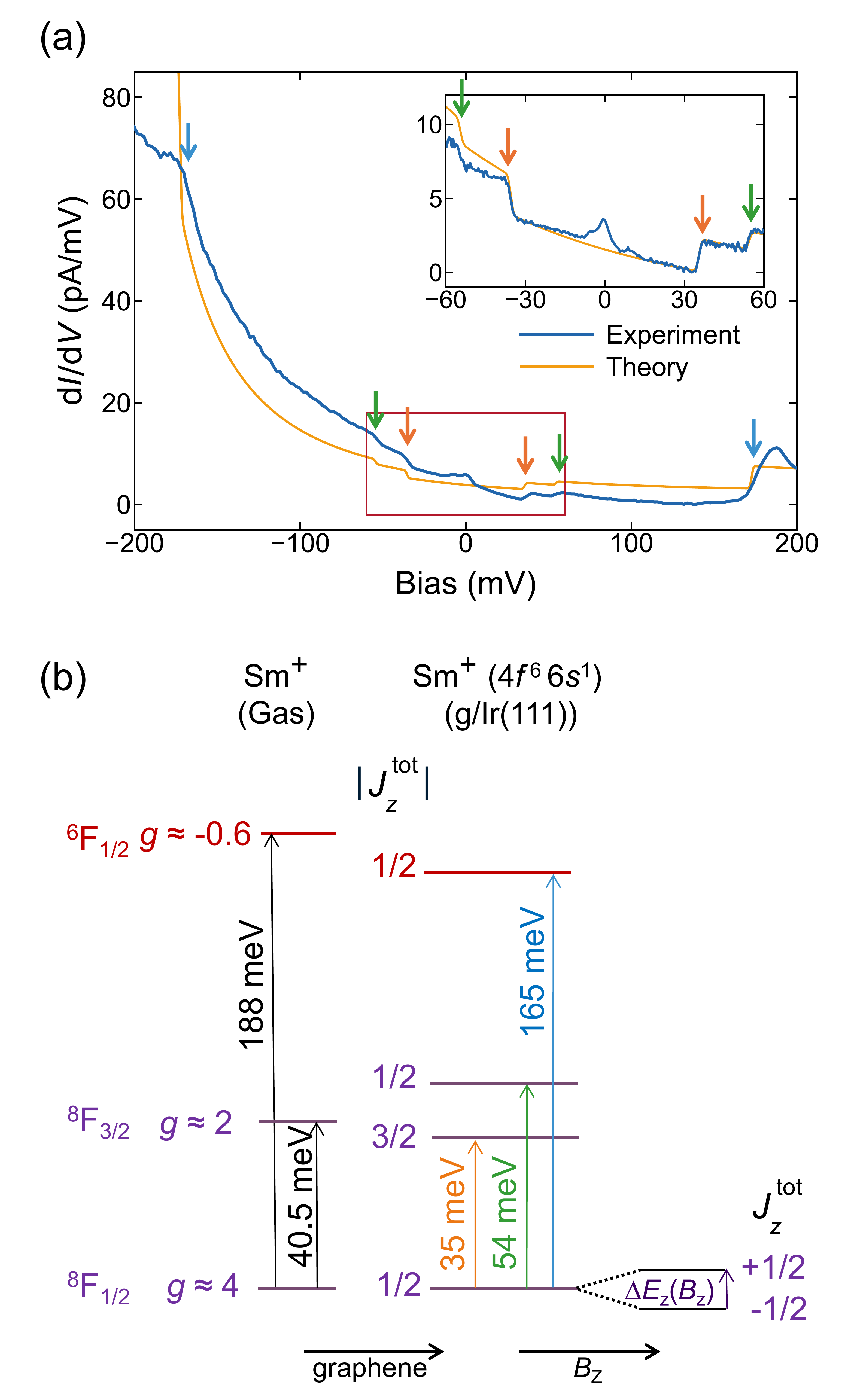}
\caption{\label{iets}(a) Main figure: Wide bias voltage range d$I$/d$V$-curve recorded on a Sm$^+$ adatom on g/Ir(111) showing three conductance steps indicated by blue, green, and yellow arrows ($V_{\rm t} = 200$~mV, $I_{\rm t} = 400$~pA). The yellow curve represents the theoretical calculation based on cotunneling theory, modeling the inelastic transitions within the adatom's electronic shells. Inset: d$I$/d$V$-curve of a smaller energy interval showing the two low energy steps more clearly ($V_{\rm t} = 50$~mV, $I_{\rm t} = 500$~pA). The curves have a parabolic, respectively, linear background that has not been subtracted. (b) Energy level diagram of Sm$^+$ in the gas-phase (left) and on g/Ir(111) (right). Colored arrows indicate the transitions giving rise to the respective conductance steps.}
\end{figure}

Figure~\ref{iets}(a) shows a wide-range d$I$/d$V$ curve of Sm on graphene/Ir(111) displaying spin excitations at $\pm 165$, $\pm 54$, and $\pm 35$~meV. The highest energy excitation was reported before and is related to intra-atomic exchange between 4$f$ and 6$s$ spins~\cite{piv20}. The two low-energy excitations are more prominent in the inset, showing a d$I$/d$V$ curve over a smaller energy range. Their origin has not been discussed, but they were already present, albeit with lower intensity, in the spectra reported in ref.~\cite{piv20}.

To identify their origin we show on the left hand side in Fig.~\ref{iets}(b) a magnetic level scheme derived from optical spectroscopy of Sm$^+$ ions in the gas phase~\cite{nist, mar78} and on the right hand side the scheme we propose for the surface adsorbed species. The two gas-phase excitation energies compare well with the ones observed here. All other tabulated charge states of Sm exhibit magnetic excitation energies far off the ones we observe. We conclude that the charge state of those surface adsorbed Sm atoms showing the magnetic excitations is very close to, if not exactly, Sm$^+$. 

The term scheme on the very left shows the total values of $S$, $L$, and $J$ in the notation $^{(2 S + 1)} {L}_{J}$, where every $L$ value is represented by a capital character, e.g., $L = 3$ is represented by F. As mentioned in the introduction, Sm preserves its gas-phase $4f^6$ filling yielding $L = 3$ and $S_{4f} = 3$. Since the 4$f$ shell is less than half-filled, both are fully anti-parallel, giving $J_{4f} = 0$. The valence spin is $S_{\rm val} = 1/2$. In the ground state, it is parallel to $S_{4f} = 3$ leading to $S = 7/2$. Since $J_{4f} = 0$, the only total angular momentum comes from the valence spin, $J = 1/2$. The term notation of the ground state is therefore $^8 {\rm F}_{1/2}$. The high-$E$ excitation flips $S_{4f}$ and $S_{\rm val}$ to anti-parallel giving rise to $S = 5/2$, thus $^6 {\rm F}_{1/2}$ in term notation. Note that $L$ and $S_{4f}$ are still fully anti-parallel and therefore $J = 1/2$ also in this excited state. The excitation energy on the surface is with 165~meV slightly lower than the gas-phase value of 188~meV. This is consistent with the 6-fold crystal field (CF) of the graphene binding site breaking the rotational symmetry. In the case of Sm on the 2-fold bridge sites of 2 ML MgO/Ag(100), this energy was 148~meV~\cite{cza25}. The lower energies in the absorbed  species are likely due to hybridization with the substrate, which induces screening of the Coulomb interaction and hence weakens the coupling between the $4f$ and $6s$ spins, thereby reducing the excitation energy.


One sees from the term notation that the low-$E$ excitation of the gas-phase cation leaves $S$ unchanged, while $J$ increases to $3/2$. Therefore it must involve a tilt between $L$ and $S_{4f}$ from fully to only partly anti-parallel such that the 4$f$-shell has a non-vanishing $J_{4f}$. On the surface, the $z$-projections $J^{\rm tot}_{z} = \pm 3/2$ and $\pm 1/2$ of this state are non-degenerate due to the magnetic anisotropy induced by the CF. We indeed observe strong out-of plane easy axis anisotropy, as the first excitation to $J^{\rm tot}_{z} = \pm 3/2$ is with 35~meV at far lower energy than the second with 54~meV. For Sm on bridge sites on MgO/Ag(100), only a single low-energy transition at 38 meV was reported~\cite{cza25}.

\subsection{Calculations of the d$I$/d$V$ spectra}
\label{sec:sts_theo}

To validate the assignment of the transitions presented in
Sec.~\ref{sec:magexc}, we quantitatively model the STS spectra using cotunneling theory \cite{del11,kyv26}, which works under the
assumption that the coupling between the adatom and the electrodes
(the STM tip and the surface) is much smaller than the charging energy
of the adatom. Only the partially-filled shells of the adatom need to
be considered. We assume that the valence electron resides
in the $6s$ shell so that the relevant part of the adatom Hamiltonian
is
$\hat{\mathcal H}_{\text{at}} = \hat{\mathcal H}_{4f}
+ \hat{\mathcal H}_{6s} + \hat{\mathcal U}_{4f4f}
+ \hat{\mathcal U}_{6s6s} + \hat{\mathcal U}_{4f6s}$, where
$\hat{\mathcal H}_{4f}$ and $\hat{\mathcal H}_{6s}$ contain the
single-electron terms of the individual shells and the operators
$\hat{\mathcal U}$ represent the Coulomb interaction within the shells
as well as between them. The $4f$-shell Hamiltonian
\begin{equation}
\label{eq:4f_shell}
\hat{\mathcal H}_{4f} = \sum_{\substack{m m'\\ \sigma \sigma'}}
 \Bigl[
 \zeta\bigl(\boldsymbol{\ell} \cdot \mathbf{s}\bigr)^{m m'}_{\sigma\sigma'}
 + \bigl(\epsilon_{4f} \delta_{m m'}
 + h_{mm'}\bigr)\delta_{\sigma \sigma '}
 \Bigr] \hat{f}^{\dagger}_{m\sigma} \hat{f}_{m'\sigma'}
\end{equation}
contains the spin-orbit coupling specified by its strength $\zeta$,
the energy of the atomic level $\epsilon_{4f}$, and the crystal-field
potential $h$ that we define using the Wybourne parameters $B_{kq}$;
in the $C_{6v}$ symmetry, there are only four non-zero crystal-field
parameters: $B_{20}$, $B_{40}$, $B_{60}$, and $B_{66}$. The $6s$-shell
Hamiltonian contains only the term involving the energy of the atomic
level, $\hat{\mathcal H}_{6s}=~\sum_{\sigma}
\epsilon_{6s}\hat{v}^{\dagger}_{\sigma} \hat{v}_{\sigma}$. The Coulomb
interaction is considered in the form of spherically symmetric
operators characterized by Slater parameters $F_k(4f,4f)$,
$k=0,2,4,6$, for the $4f$ shell, $F_0(6s,6s)$ for the $6s$ shell, and
$F_0(4f,6s)$ and $G_3(4f,6s)$ for the interaction between the shells
\cite{bookCondon}.

The interaction between an $s$ shell and a shell with an
orbital momentum quantum number $\ell$ reads as
\begin{multline}
\hat{\mathcal U}_{\ell\text{--}s}=
F_0 \sum_{m}\sum_{\sigma\sigma'}
\hat v_{\sigma'}^{\dagger}
\hat v_{\sigma'}
\hat f_{m\sigma}^{\dagger}
\hat f_{m\sigma}\\
-\frac{G_\ell}{2\ell+1} \sum_{m}\sum_{\sigma\sigma'}
\hat v_{\sigma'}^{\dagger}
\hat v_{\sigma}
\hat f_{m\sigma}^{\dagger}
\hat f_{m\sigma'}\,,
\end{multline}
which can be rewritten into an equivalent form
\begin{equation}
\hat{\mathcal U}_{\ell\text{--}s}=
\biggl(F_0-\frac{G_\ell}{2(2\ell+1)}\biggr) \hat N_{6s}\hat N_{4f}
- \frac{2G_\ell}{2\ell+1}\hat{\mathbf S}_{6s}\cdot\hat{\mathbf S}_{4f}
\label{eq:Coul_sf}
\end{equation}
that expresses the interaction in terms of the shell occupation
numbers $\hat N$ and of the spins $\hat{\mathbf S}$ carried by the
shells. The second term in Eq.~\eqref{eq:Coul_sf} is a ferromagnetic
Heisenberg exchange with exchange constant
$J_{\rm H}=G_{\ell}/(2\ell+1)$, which links the current model to the
discussion in \cite{piv20}. A general Coulomb interaction between
shells with orbital momentum quantum numbers $\ell$ and $\ell'$, such
as between $4f$ and $5d$, contains more terms involving also the
orbital momenta $\hat{\mathbf L}$ carried by the shells.

\begin{figure}
\includegraphics[width=\linewidth]{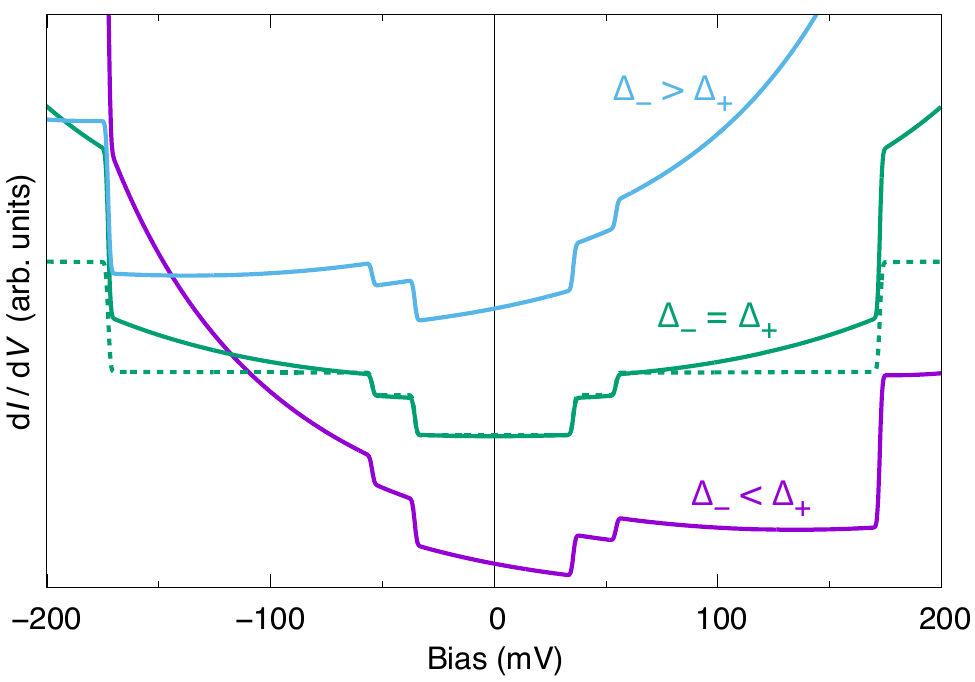}
\caption{\label{fig:param6s}Dependence of the theoretical d$I/$d$V$ spectrum
on the parameters of the Sm 6s shell. The spectrum is symmetric with
respect to the bias polarity when the charging energies for removal,
$\Delta_-$, and addition, $\Delta_+$, of an electron to the 6s shell
are equal (green, $\epsilon_{6s}=-0.33$~eV). When it is easier to
remove the electron ($\Delta_-<\Delta_+$, violet,
$\epsilon_{6s}=~-0.2$~eV), the spectrum increases faster at negative
biases, and when it is easier to add the electron ($\Delta_->\Delta_+$,
light blue, $\epsilon_{6s}=-0.46$~eV), the spectrum increases faster at
positive biases. The spectra plotted with solid lines are computed with
$F_0(6s,6s)=0.8$~eV, the effect of an increased Coulomb repulsion is
illustrated for the symmetric spectrum with a dashed line ($F_0(6s,6s)=8$~eV,
$\epsilon_{6s}=-3.93$~eV).}
\end{figure}

Finally, we need to define how the adatom couples to the STM tip
(hosting conduction states $\hat{a}^\dagger_{k\sigma}$) and to the surface
(hosting conduction states $\hat{b}^\dagger_{k\sigma}$). Since the
$4f$ shell is very compact, we treat it as completely decoupled from
the electrodes and consider only tunneling between the electrodes and
the diffuse $6s$ shell. We choose the tunneling operator in the form
\begin{equation}
\hat{\mathcal{V}} = A \sum_{k\sigma}
\hat{a}^\dagger_{k\sigma}\hat{v}_{\sigma}
+ B\sum_{k\sigma} \hat{b}^{\dagger}_{k\sigma}  \hat{v}_{\sigma} 
+ \text{h.c.}
\label{eq:tunneling}	
\end{equation}
with the tunneling amplitudes $A$ and $B$ independent of the $k$
vector. Similarly, we assume the densities of states of
the tip and of the surface to be constant, which is reasonable in the small energy range around the Fermi level that is probed in the STS measurements.

The numerical values of the parameters entering the Hamiltonian
$\hat{\mathcal H}_\text{at}$ are determined as follows: The Slater
parameters $F_k$ for $k\geq 2$ are computed for the
Sm\textsuperscript{+} ion in the $4f^6 6s^1$ configuration with the
Cowan code \cite{bookCowan} and then reduced to 80\% of their computed
values, following the strategy often used when ionic models are
employed to interpret core-level spectroscopies \cite{oga94}. The parameters
$F_0$, $\epsilon_{4f}$, $\epsilon_{6s}$ need to be chosen so that
the $4f^6 6s^1$ configuration has the lowest energy among all $4f^n 6s^m$
configurations.

Once the correct ground-state configuration is fixed, the
$4f$-shell parameters $F_0(4f,4f)$ and $\epsilon_{4f}$ do not
influence the shape of the STS spectrum any more since the $4f$ shell does
not participate in the tunneling and its occupation remains constant. The
$6s$-shell parameters affect the spectrum as illustrated in
Fig.~\ref{fig:param6s}. For a fixed $F_0(6s,6s)$, the charging
energies corresponding to addition of a $6s$ electron, $\Delta_+=E(4f^6
6s^2)-E(4f^6 6s^1)$, and removal of a $6s$ electron, $\Delta_-=E(4f^6
6s^0)-E(4f^6 6s^1)$, are controlled by $\epsilon_{6s}$. When removing
the electron is easier ($\Delta_-<\Delta_+$), the differential
conductance d$I$/d$V$ increases faster at negative biases, and when
adding the electron is easier ($\Delta_->\Delta_+$), then d$I$/d$V$
increases faster at positive biases. The increase is more non-linear
the closer the voltage bias~$V$ is to $\Delta_+$ on the positive side
or to $-\Delta_-$ on the negative side. At these voltages, our theory
diverges, which is an artifact of the perturbative nature of the
employed approximation \cite{kyv26} -- in reality, the differential
conductance stays finite. The voltage interval between the divergences
grows with $F_0(6s,6s)$, namely $\Delta_- + \Delta_+ \approx 2
F_0(6s,6s)$.

The measured spectra that display the high-energy spin
excitations (Fig.~\ref{iets}a) increase faster at negative biases and
are well reproduced with $F_0(6s,6s)=0.8$~eV and
$\epsilon_{6s}=~-0.11$~eV, which gives $\Delta_-=0.24$~eV and
$\Delta_+=~0.67$~eV. The limited stability toward losing the 6s
electron is consistent with the charge state changes discussed in
Secs.~\ref{sec:charge_and_dist} and~\ref{sec:charge_and_magexc}. The
STS spectra of the alternate charge state (Fig.~\ref{man}b) show a
faster growth at positive biases, which indicates limited stability
toward capturing an electron to the 6$s$ shell and hence the charge
state is likely Sm\textsuperscript{2+} in the $4f^6 6s^0$ configuration.

The remaining parameters of our adatom Hamiltonian are fitted to
reproduce the
energies of the observed magnetic excitations; $G_3(4f,6s)=165.5$~meV
controls the high-energy excitation, and $\zeta=134.7$~meV and
$B_{20}=72$~meV recover the two low-energy excitations. The
other $B_{kq}$ are set to zero, since we do not have enough
experimental information to reliably determine them. The chosen value of the
spin-orbit parameter $\zeta$ corresponds to 96\% of the Cowan-code
estimate for the Sm\textsuperscript{+} ($4f^6 6s^1$) ion.

\begin{figure}
\includegraphics[width=0.65\linewidth]{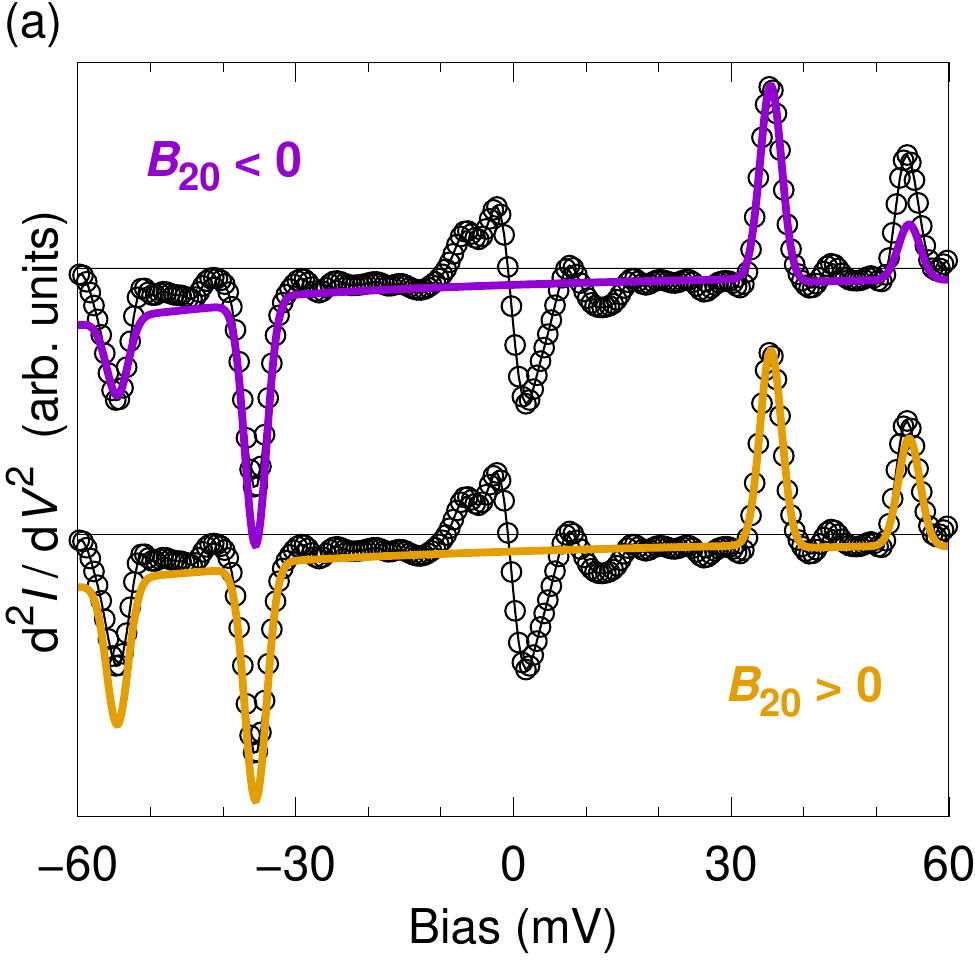}%
\includegraphics[width=0.35\linewidth]{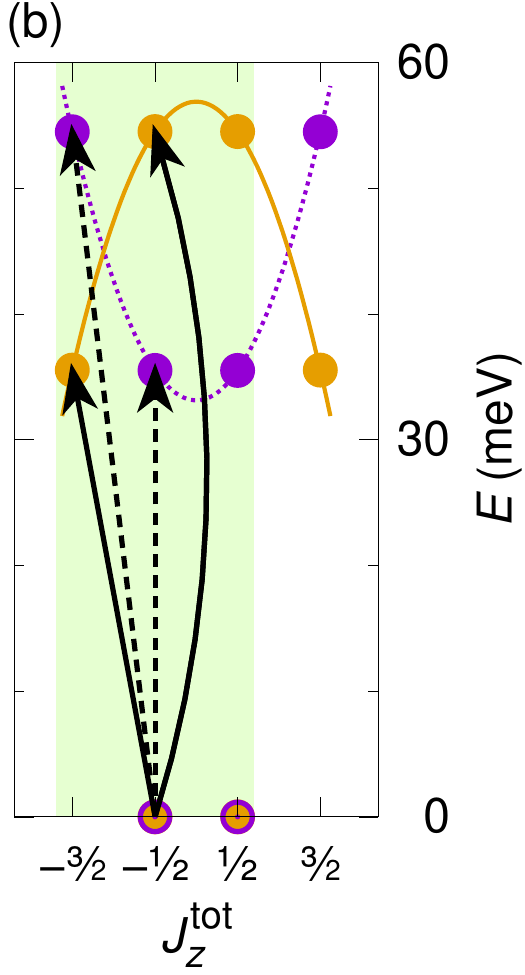}%
\caption{\label{fig:d2IdV2}(a) d$^2I$/d$V^2$ curve corresponding to
  the d$I$/d$V$ curve shown in the inset of Fig.~\ref{iets}(a) -- the
  experiment (black symbols), a theory assuming $B_{20}>0$ (orange
  line), and an alternative theory assuming $B_{20}<0$ (violet
  line). (b) Detail of the transitions from the $^8 {\rm F}_{1/2}$
  ground state doublet to the crystal-field split $^8 {\rm F}_{3/2}$
  quadruplet for the two signs of $B_{20}$. All transitions fulfilling
  $\Delta J_z^{\rm tot} = 0, \pm 1$ are allowed (all states in the
  light green rectangle are accessible from the $J_z^{\rm tot}=-1/2$
  component of the ground state), albeit some of them with very small
  probabilities. The dominant transitions are indicated by arrows
  (solid line for $B_{20}>0$, dashed line for $B_{20}<0$).}
\end{figure}

The STS spectra computed by the method detailed in \cite{kyv26} are
shown in Figs.~\ref{iets}(a) and~\ref{fig:d2IdV2}(a) together with the
experimental data. The alignment of the theory to experiment is done in two
steps. First, an overall scaling factor is adjusted to match the
intensity of the d${}^2I/$d$V^2$ peak at the positive bias corresponding
to the first excitation, Fig.~\ref{fig:d2IdV2}(a). This can be viewed
as setting the tunneling amplitudes $A$
and $B$ in Eq.~\eqref{eq:tunneling} since the cotunneling differential
conductance is proportional to $|A B|^2$. Second, the d$I$/d$V$ theory
is vertically shifted to match the d$I$/d$V$ data at biases smaller
than the first excitation ($|V|\lesssim 30$~meV),
Fig.~\ref{iets}(a). This corresponds to adding an appropriate elastic
contribution to the computed spectra, which a theory that considers
tunneling through just the Sm $6s$ shell cannot fully capture.

As described above, the energies of the three inelastic excitations
are fitted (they are an input to our theory). The
heights of the steps in the differential conductance (or the intensities
of the peaks in the d${}^2I/$d$V^2$ spectrum), are a result of the
theory. We see that the relative d${}^2I/$d$V^2$
intensity of the low-energy excitations comes out very well, the
relative intensity of the high-energy excitation with respect to the
low-energy excitations appears underestimated.

The origin of the main transitions observed in the STS spectrum is
indicated in Fig.~\ref{fig:d2IdV2}(b). Since the electrons tunnel
only through the $6s$ shell that does not have any orbital degrees of
freedom, the tunneling electrons can exchange only spin with the
adatom. This, together with the fact that our Hamiltonian
$\hat{\mathcal H}_\text{at}$ commutes with $\hat J_z^\text{tot}$ (as
long as we keep $B_{66}=0$) implies that only transitions
fulfilling the condition $\Delta J_z^{\rm tot} = 0, \pm 1$ are
accessible in STS \cite{kyv26}. There are many such transitions but
only a few of them have a large enough probability to be visible in the
measurements reported in this paper and those correspond to the
transitions suggested in Sec.~\ref{sec:magexc}.

The calculations discussed so far assumed a positive $B_{20}$ but the
two low-energy excitations can be reproduced also with a negative
$B_{20}$ when the values $\zeta=130.9$~meV and $B_{20}=-80.4$~meV are
utilized as illustrated in Fig.~\ref{fig:d2IdV2}(a). This set of
parameters, however, does not reproduce the experimental data as
accurately as the set with a positive $B_{20}$. Namely, the ratio of
intensities of the second and the first transition comes out noticeably
smaller than in the measurements. We therefore infer that $B_{20}$ is indeed
positive in the Sm adatom.


\subsection{Kondo resonance}
\begin{figure}[ht!]
\includegraphics[width = \linewidth]{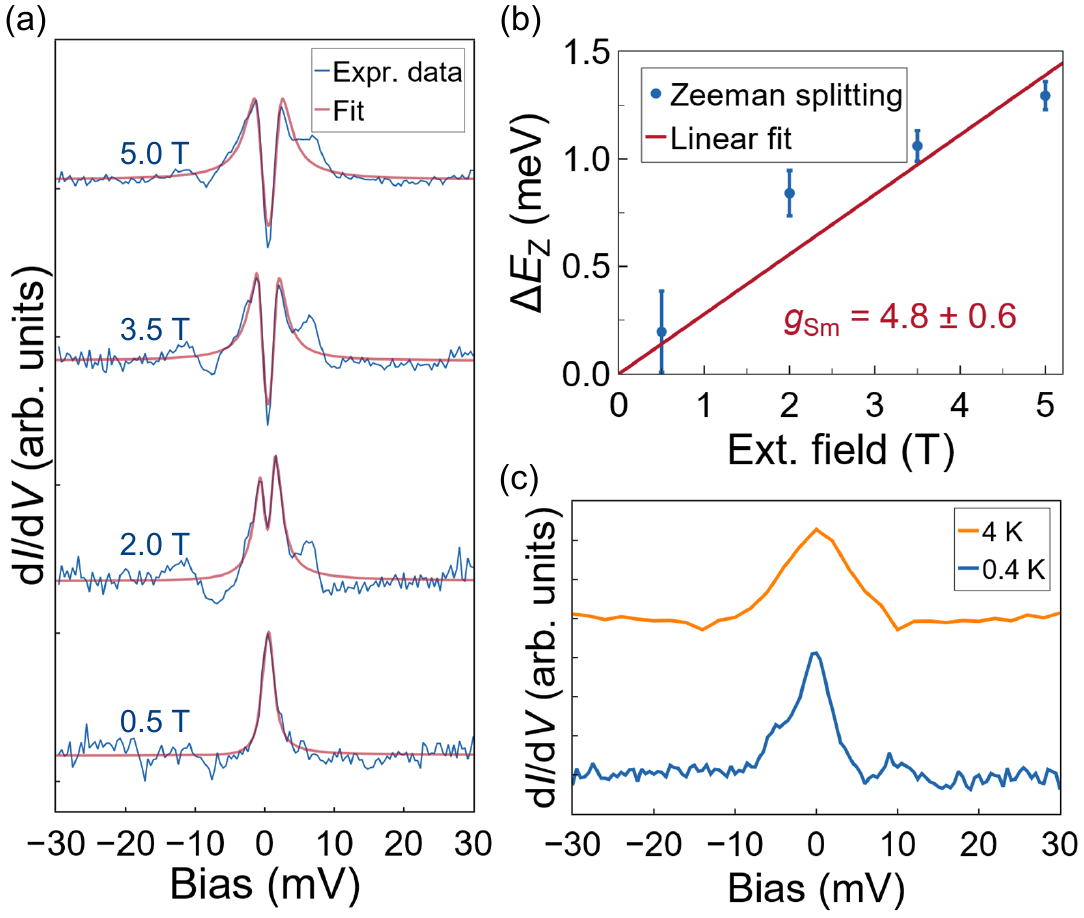}
\caption{(a) d$I$/d$V$-curves showing Kondo peak and its magnetic field splitting for individual Sm atoms adsorbed on g/Ir(111). Blue curves are substrate background corrected experimental data and red ones are fits ($V_{\rm t} =~40$~mV, $I_{\rm t} = 400$~pA, {\bf B} perpendicular to the surface). (b) Zeeman splitting $\Delta E_{\rm Z}$. (c) Background-corrected d$I$/d$V$ spectra acquired at 4~K ($V_{\rm t} = 200$~mV, $I_{\rm t} = 400$~pA) and 0.4~K ($V_{\rm t} = 50$~mV, $I_{\rm t} = 500$~pA).}
\label{kon}
\end{figure}

In addition to the inelastic conductance steps reported above, we observe a sharp peak centered at the Fermi level, see 0.5~T spectrum in Fig.~\ref{kon}(a). This peak is also present without an external magnetic field, see Fig.~\ref{fit}(a), it shows Zeeman splitting, see Figs.~\ref{kon}(a) and (b), and it broadens with temperature beyond what is expected from Fermi-Dirac broadening, see Fig.~\ref{kon}(c). Taken together, these observations identify this feature as Kondo peak, indicating that Sm is strongly coupled to graphene and/or the underlying metal substrate. 

As seen from inspection of Fig.~\ref{kon}(a) for $B \ge 3.5$~T, each of the Zeeman split peaks exhibits an asymmetric profile, with a much steeper drop-off towards 0~mV. This is attributed to inelastic spin-flip excitations between the field-split $-1/2$ and $+1/2$ of the ground state doublet, see right hand side of Fig.~\ref{kon}(b)~\cite{ott08, zha13}. By correcting the background and fitting the spectra with a sum of two Fano functions (see details in Appendix~\ref{app:Kondo-fit}), we extract the Zeeman splittings shown in Fig.~\ref{kon}(b). The linear fit is not perfect but it gives an estimate of $g = 4.8 \pm 0.6$ and thereby confirms the very large $g$-factor of the adatoms. The value is close to $g = 3.95$ of gas-phase $\rm Sm^+$ and in agreement with $g \sim 5$ of Sm${^+}$ on MgO/Ag(100) substrate~\cite{mar78, nist, cza25}. 

Figure~\ref{kon}(c) shows that the peak-width increases with temperature. This is due to a combined effect of Fermi-Dirac broadening of the tip and the intrinsic temperature behavior of the Kondo system~\cite{ott08}. Fermi liquid theory gives the intrinsic full width at half-maximum of the Kondo resonance as $2 \, \Gamma = \sqrt{(\alpha k_{\rm B} T)^2 + (2 k_{\rm B} T_{\rm K})^2}$~\cite{nag02, ott08, zha13}, where $T_{\rm K}$ is the Kondo temperature, $\alpha$ the asymptotic slope at $T \gg T_{\rm K}$, and $k_{\rm B}$ the Boltzmann constant. From the two data points we have for the intrinsic broadening, $2 \, \Gamma (0.4 \, {\rm K}) = 3.2$~meV and $2 \, \Gamma (4 \, {\rm K}) = 4.9$~meV, we estimate $T_{\rm K} = 18.6 \pm 1.8$~K. Since all measurements performed with our 0.4~K STM need to be taken at $B \ge 0.5$~T, we report in Fig.~\ref{fit}(a) in Appendix~\ref{app:Kondo-fit} a d$I$/d$V$-curve recorded at $B = 0$~T with a different STM setup operating at 6.3~K and equally showing a Kondo peak. Note that this spectrum also shows the first of the two magnetic low-$E$ steps. 

\section{Conclusions}
We investigated by means of low-$T$ STM the charge state, magnetic properties, as well as the Kondo features of individual Sm atoms adsorbed onto graphene/Ir(111). While isolated Sm atoms show no magnetic excitations in d$I$/d$V$-curves, such excitations become visible once the atoms have a neighbor at 1~nm distance. When more than two atoms are close, d$I$/d$V$-steps appear already at larger interatomic distances. We observe charge rings in some of the atoms that we attribute to a reversible change of charge state induced by the electric field of the STM tip. This tip-induced charging suggests that the bias applied on the STM tip acts as a gate voltage, allowing the reversible control over the charge states of adatoms. Each atom that exhibits such a charge ring also shows d$I$/d$V$-steps, but also atoms without charge ring might display such steps. From comparison with gas-phase spectroscopy, we infer that the Sm atoms are close to one-fold charged cations. Two hitherto not discussed low-energy spin excitations imply a transition from fully anti-parallel to partly anti-parallel 4$f$-orbital and spin moments and give access to the large magneto-crystalline out-of-plane anisotropy of the atoms. Comparison with theory confirms Sm$^+$ as the ground state, reveals Sm$^{2+}$ as being energetically close, and therefore very likely the second charge state observed in experiment. The magnetic excitations in d$I$/d$V$ spectra are very well reproduced and give access to crystal field parameters of the $4f$ shell and, by extension, of the excited $J=3/2$ quadruplet. We observe a Kondo resonance and infer from its field-splitting the large $g$-factor of the adatoms.

\section*{Acknowledgments}
We acknowledge support from the Swiss National Science Foundation (TMAG-2\_209266) and from the Grant-in-Aid for Scientific Research (S) (No.19H05621) and (B) (No. 24K01339).
The contribution of D.~K. and J.~K. was co-funded by the European Union
and the Czech Ministry of Education, Youth and Sports (Project TERAFIT --
{\text{CZ.02.01.01/00/22\_008/0004594}}).


\section*{Data availability}
The results of the theoretical modeling will be available as a dataset
in \cite{data_theo}.

\appendix

\section{Analysis of Kondo resonance}
\label{app:Kondo-fit}
\begin{figure}
\centering
\includegraphics[width = \linewidth]{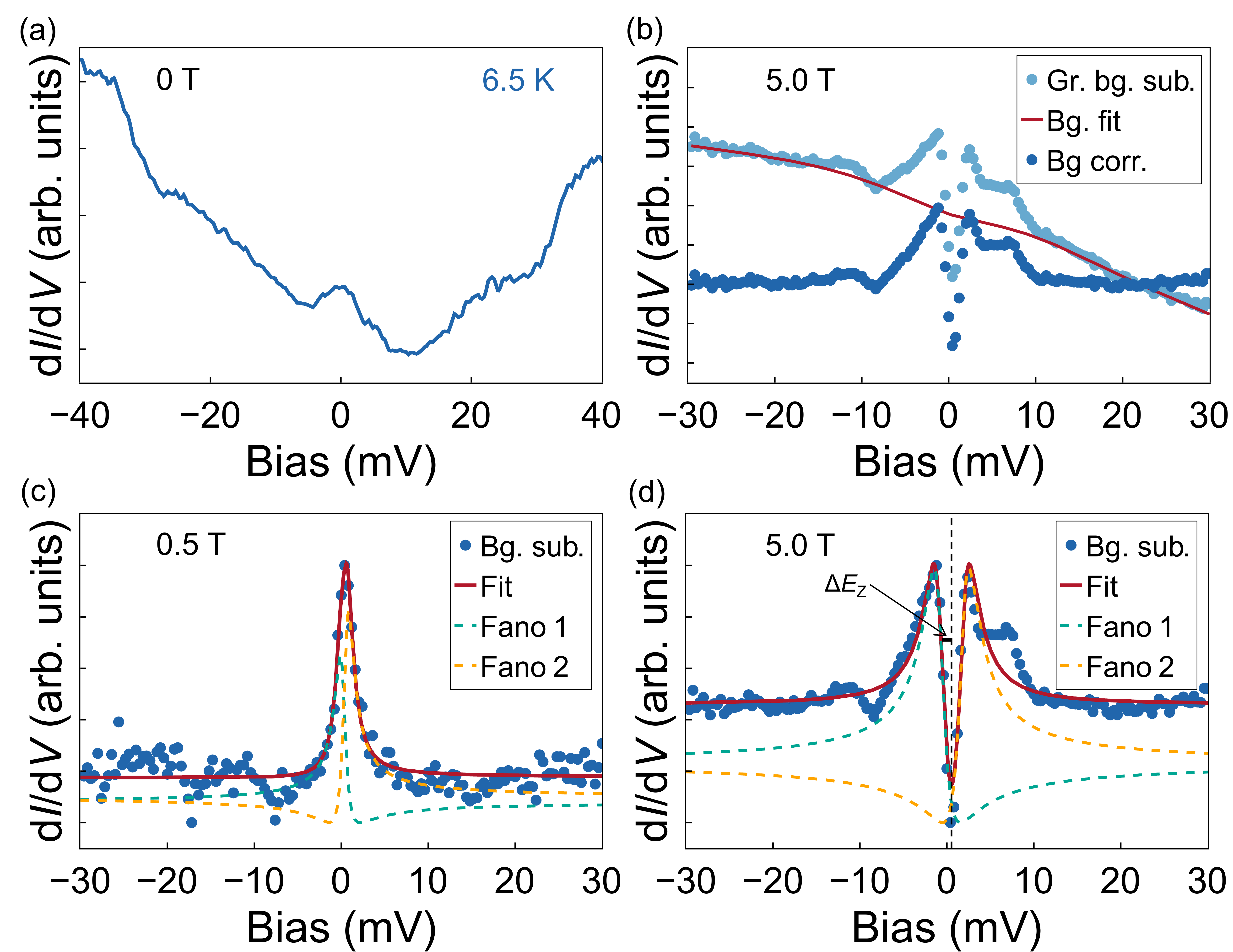}
\caption{(a) d$I$/d$V$ spectrum showing the Kondo peak of Sm in the absence of an external magnetic field ($T = 6.5$~K, $I_{\rm t} = 200$~pA, $V_{\rm t} = -200$~mV). (b) d$I$/d$V$ spectra at 5.0~T ($I_{\rm t} = 400$~pA, $V_{\rm t} = 40$~mV). The  light blue points show the graphene background subtracted data. The dark blue points show the data after removing slope background with ALS baseline correction. The solid red line shows the baseline obtained from the ALS method. (c) and (d) Background-corrected spectrum measured at 0.5~T and 5.0~T ($I_{\rm t} = 400$~pA, $V_{\rm t} = 40$~mV). The dashed lines represent the two Fano line shapes used to fit the Kondo resonance. The red solid line represents the sum of the two Fano line shapes, and fits the data quite well. The Zeeman splitting $\Delta E_{\rm Z}$ is obtained from the two Fano fits in Eq.~\eqref{eq:model}. In (d) $\Delta E_{\rm Z}$ is indicated by the black arrow.}
\label{fit}
\end{figure}

A two-step background subtraction and a Fano fitting were performed on the magnetic-field-dependent spectra to analyze the Zeeman splitting of the Kondo peak. Taking the Sm spectrum at 5~T as an example, we first subtracted the graphene background, resulting in the light blue $\mathrm{d}I/\mathrm{d}V$-curve shown in Fig.~\ref{fit}(b). This spectrum still shows a non-linear slope intrinsic to the background. To address this, we employed a baseline subtraction method based on asymmetric least squares (ALS). The resulting background-corrected spectrum is shown in dark blue in Fig.~\ref{fit}(b)--(d). For all spectra acquired in Fig.~\ref{kon}(a), the same ALS parameters were applied to ensure consistency. While some additional peaks remained after the data processing, they showed no dependence on the applied magnetic field and did not interfere with the analysis.

The Kondo resonance was fitted with two Fano line shapes (Figs.~\ref{fit}(c) and (d))~\cite{ter08, liu15}:
\begin{equation}
    \frac{\mathrm{d} I}{\mathrm{d} V}(V) = F_1(V) + F_2(V)
    \label{eq:model}
\end{equation}
where $F_1(V)$ and $F_2(V)$ are given by:
\begin{equation}
    F_{1,2}(V) = \frac{a_{1,2}(\epsilon_{1,2}+q)^2}{1+\epsilon_{1,2}^2}
    \label{eq:fano}
\end{equation}
where $\epsilon_{1,2} = e(V - V_{1,2})/\Gamma$ is the normalized energy, $a_{1,2}$ are the amplitude of the Fano line shapes, $\Delta E_{\rm Z}~=~\left| e(V_1-V_2)/2 \right|$ is the Zeeman splitting energy, $\Gamma$ is the half width at half maximum, $q$ is the Fano factor, and $e$ is the elementary charge. $V_{1,2}$, $a_{1,2}$, $\Gamma$, and $q$ are fit parameters. 

\section{Charge ring for dysprosium adatoms}
\begin{figure}[!htbp]
\centering
\includegraphics[width = .25\textwidth]{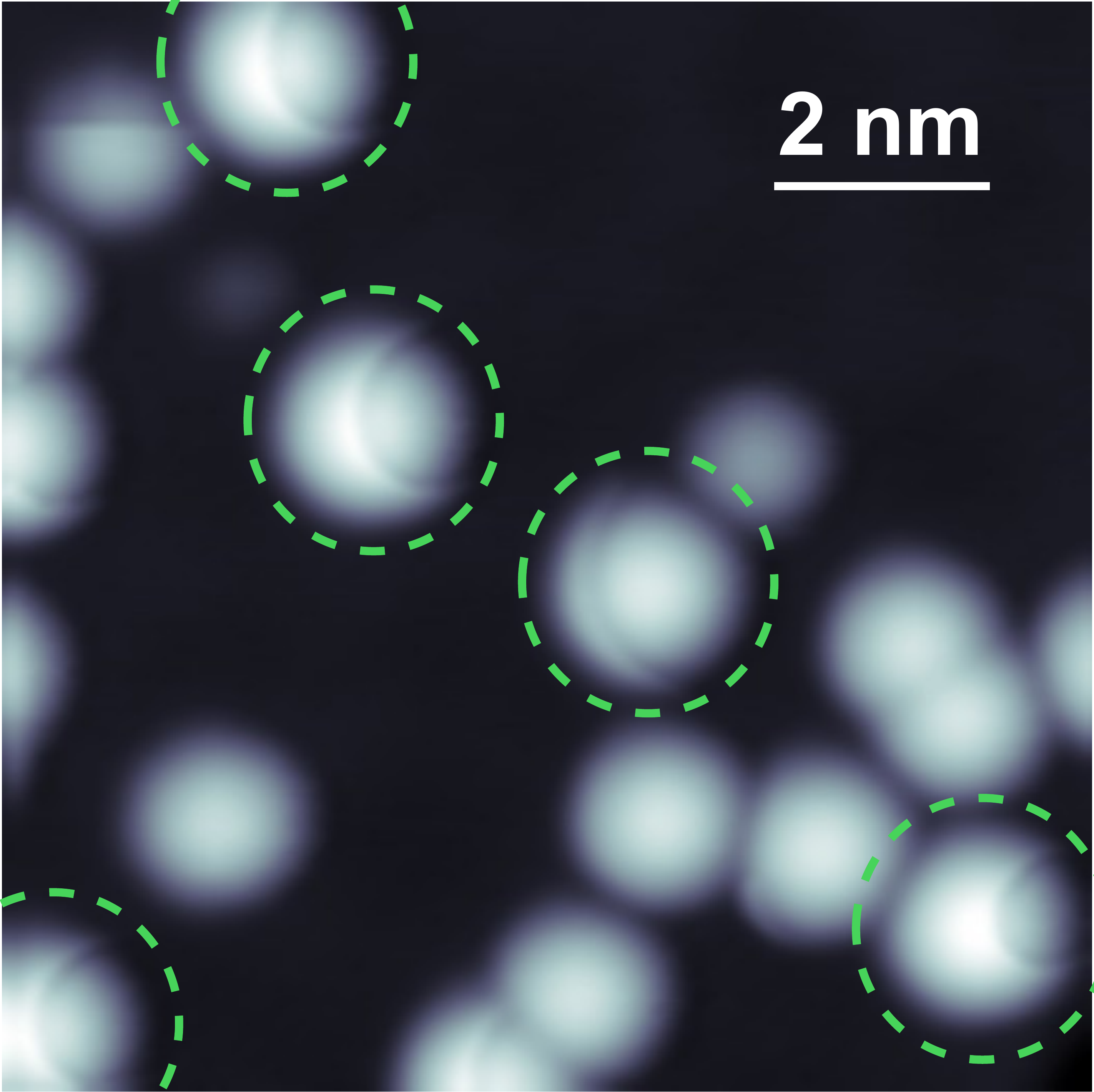}
\caption{STM topography of Dy adatoms on g/Ir(111) ($\Theta_{\rm Dy} = 8 \times 10^{-3}$~ML, $V_{\rm t} =~100$~mV, $I_{\rm t} = 10$~pA). Green dashed circles highlight adatoms displaying charge rings.}
\label{Dy}
\end{figure}

Figure~\ref{Dy} shows a representative STM image of Dy adatoms on graphene/Ir(111). Several atoms exhibit charge rings, as highlighted by green dashed circles. In contrast to Sm, Dy adatoms without nearby neighbors can also display charge rings.

\section{Sm\textsuperscript{+} level diagram and the transitions allowed
  in STS in more detail}
\label{app:Sm-g-Cu}

The level diagram shown in Fig.~\ref{iets}(b) and discussed in
Sec.~\ref{sec:magexc} does not include all levels of the
Sm\textsuperscript{+} ion with energies below 200~meV, it shows only
those that have an appreciable probability to be excited by inelastic
tunneling. Here we provide additional details in the language of the
model introduced in Sec.~\ref{sec:sts_theo}.

\begin{figure}
\includegraphics[width=0.6\linewidth]{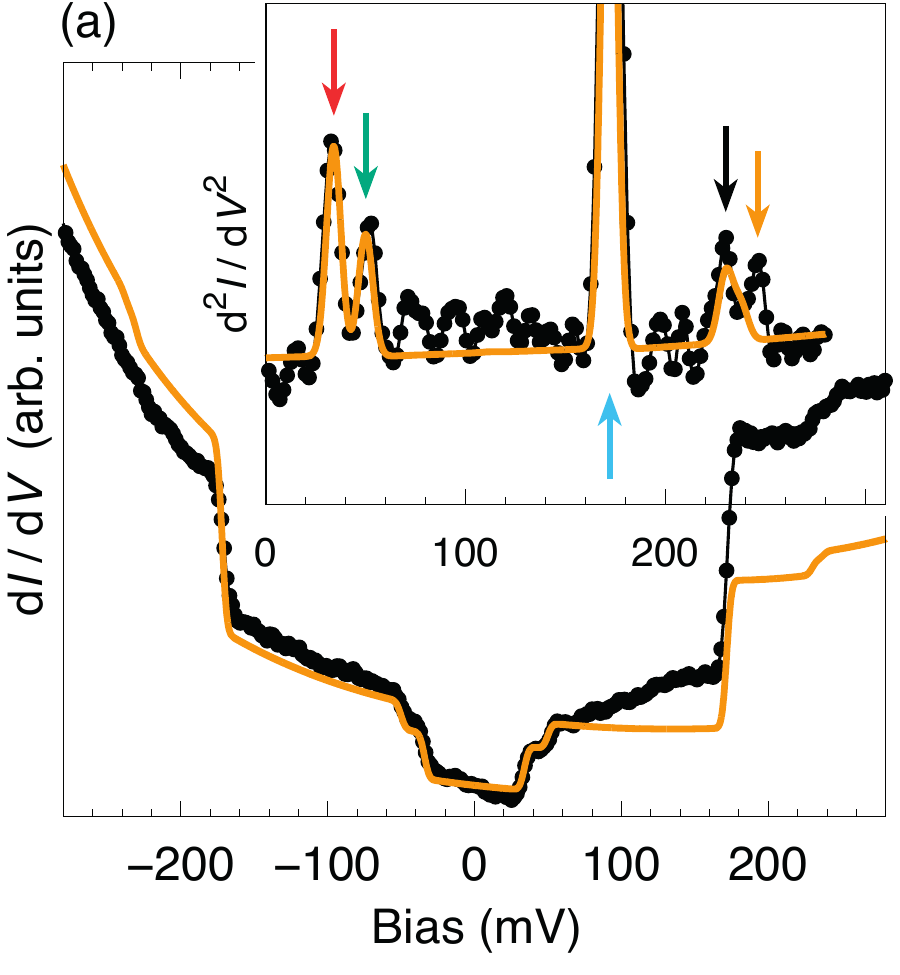}%
\includegraphics[width=0.4\linewidth]{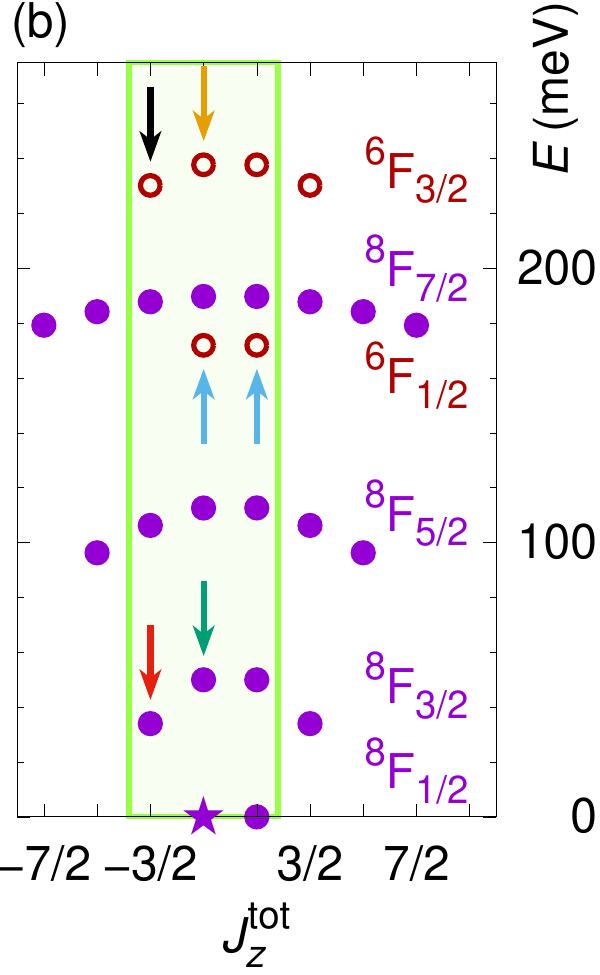}%
\caption{\label{fig:Sm-g-Cu}(a) Measured d$I$/d$V$ spectrum (black
  symbols) compared with calculations (yellow curves). The
  calculated spectrum has a Gaussian broadening of 6~mV FWHM added. The
  experimental data correspond to Sm on g/Cu from \cite{piv20}
  ($V_{\rm t} = -300$~mV, $I_{\rm t} = 200$~pA, $V_{\rm mod}
    = 5$~mV, $T = 4.3$~K). The inset shows the numerical derivative
  $\text{d}^2I/\text{d}V^2$ at positive biases; the corresponding
  theoretical spectrum has a Gaussian
  broadening of 8.5~mV FWHM added (the numerical derivative involved
  an additional smoothening). (b) All transitions fulfilling $\Delta
  J_z^{\rm tot} = 0, \pm 1$ are allowed, that is, all states in the
  light green rectangle are accessible from the $J_z^{\rm tot}=-1/2$
  component of the ground state (marked with a star). The main
  transitions discernible in (a) are indicated by matching arrows in
  both panels.}
\end{figure}

The spins of the $4f$ and $6s$ shells, $S_{4f}=3$ and $S_{6s}=1/2$,
are coupled by ferromagnetic exchange, which splits the
$S_{4f}\otimes S_{6s}$ manifold into states with total spins $S=7/2$
(parallel $S_{4f}$ and $S_{6s}$) and $S=5/2$ (antiparallel $S_{4f}$ and
$S_{6s}$) separated by approximately 165~meV. The spin-orbit
interaction couples these spins with the orbital momentum of the $4f$
shell, $L_{4f}=3$, which further splits each of the two $S\otimes
L_{4f}$ manifolds into a series of states with total momenta
$J=1/2,3/2,\dots,13/2$ and $J=1/2,3/2,\dots,11/2$
(${}^8\text{F}_{1/2},{}^8\text{F}_{3/2},\dots,{}^8\text{F}_{13/2}$
and ${}^6\text{F}_{1/2},{}^6\text{F}_{3/2},\dots,{}^6\text{F}_{11/2}$
in the spectroscopic notation). The gaps between these terms are
smaller than the splitting due to the exchange between the $4f$ and $6s$
spins. Each of the terms is then further split by the crystal-field
potential. The energy spectrum of the ion up to 250~meV is plotted in
Fig.~\ref{fig:Sm-g-Cu}(b) for parameters corresponding to Sm on
graphene/Cu as reported in \cite{piv20}: $F_0(6s,6s)=1.3$~eV,
$\epsilon_{6s}=-0.45$~eV, $G_3(4f,6s)=165.5$~meV,
$\zeta=129$~meV and $B_{20}=60.5$~meV. The other $B_{kq}$ are set to
zero as for Sm on graphene/Ir(111). We shift our discussion to
the graphene/Cu substrate because we have STS data with a good
resolution up to higher voltage biases than for the graphene/Ir(111)
system discussed in the rest of the paper.

The STS spectrum is shown in Fig.~\ref{fig:Sm-g-Cu}(a) in comparison
with model calculations. Apart from
the transitions discussed in Secs.~\ref{sec:magexc}
and~\ref{sec:sts_theo}, there is also a weak transition to the
${}^6\text{F}_{3/2}$ multiplet predicted by the theory, a hint of
which seems to be present also in the experimental data. The
${}^8\text{F}_{5/2}$ and ${}^8\text{F}_{7/2}$ multiplets also fall
within the range probed by the experiment, but their transition
probabilities are too small to be visible (the transition to
${}^8\text{F}_{5/2}$ is three orders and to ${}^8\text{F}_{7/2}$ four
orders of magnitude less intense than the spin-flip transition to
${}^6\text{F}_{1/2}$).

The classification of the energy levels using the term symbols
neglects coupling to higher multiplets in the individual shells and
assumes the crystal-field splitting being much smaller than the
spin-orbit coupling. Our numerical calculations make neither of these
assumptions. If we, however, restricted the Hilbert space to the
$L_{4f}\otimes~ S_{4f}\otimes~ S_{6s}$ manifold and made the
crystal-field splitting infinitesimally small, the term symbols would
become exact. Furthermore, for large Coulomb repulsion, the STS
transition probabilities from a state $|\psi\rangle$ to a state
$|\psi'\rangle$ would become proportional to \cite{del11}
\begin{equation}
\sum_{\alpha\in\{x,y,z\}}
 \bigl|\langle\psi'|\hat S^{6s}_\alpha|\psi\rangle\bigr|^2\,.
\end{equation}
These matrix elements of the $6s$ spin operator can be evaluated with
the help of the Wigner--Eckart theorem. The spin is a (spherical)
vector operator and hence we get selection rules analogous to the x-ray
absorption in the dipole approximation; the transition from
$|J,J_z\rangle$ to $|J',J'_z\rangle$ is allowed only if $|J'-J|\leq 1$
and $|J'_z-J_z|\leq 1$. For the Sm\textsuperscript{+} adatom, only
transitions from the ${}^8\text{F}_{1/2}$ ground state to the
${}^8\text{F}_{3/2}$, ${}^6\text{F}_{1/2}$ and ${}^6\text{F}_{3/2}$
excited states are allowed in this approximation.


\newpage
\bibliography{ms_10}
\end{document}